\documentclass[finnish,oneside,openright,utf]{HYgradu}

\usepackage{lmodern} 
\usepackage{textcomp} 
\usepackage[pdftex]{color, graphicx} 
\usepackage{fancyhdr} 
\usepackage{tikz} 
\usepackage{amsmath, amssymb} 
\usepackage[footnotesize,bf]{caption} 
\usepackage{blindtext}
\usepackage{titlesec}
\usepackage[titletoc]{appendix}
\usepackage{physics}
\usepackage{biblatex}
\usepackage[pdftex, plainpages=false]{hyperref} 

\usepackage{cancel}
\usepackage{listings}

\definecolor{mygreen}{rgb}{0,0.6,0}
\definecolor{mygray}{rgb}{0.5,0.5,0.5}
\definecolor{mymauve}{rgb}{0.58,0,0.82}
\title{Mukautuva hilantarkennus Vlasiator-plasmasimulaattorissa}
\author{Leo Kotipalo}
\date{24.11.2021}
\level{Kandidaatintutkielma}
\subject{Teoreettinen fysiikka}
\faculty{Matemaattis-luonnontieteellinen tiedekunta}
\programme{Fysikaalisten tieteiden kandidaattiohjelma}
\address{PL 64 (Gustaf Hällströmin katu 2a)\\00014 Helsingin yliopisto}
\prof{Yann Pfau-Kempf}
\censors{Yann Pfau-Kempf}{Markus Battarbee}{}
\keywords{Avaruusplasmafysiikka, AMR, numeerinen mallinnus}
\depositeplace{}
\additionalinformation{}
\classification{}

\hypersetup{
    bookmarks=true,         
    unicode=true,           
    pdftoolbar=true,        
    pdfmenubar=true,        
    pdffitwindow=false,     
    pdfstartview={FitH},    
    pdftitle={},            
    pdfauthor={},           
    pdfsubject={},          
    pdfcreator={},          
    pdfproducer={pdfLaTeX}, 
    pdfkeywords={something} {something else}, 
    pdfnewwindow=true,      
    colorlinks=true,        
    linkcolor=black,        
    citecolor=black,        
    filecolor=magenta,      
    urlcolor=cyan           
}

\begin{document}
\maketitle


\begin{abstract}
Avaruusplasman simulointi globaalissa mittakaavassa on laskennallisesti vaativaa.
Mallintamalla eri alueita eri tarkkuudella laskentavaatimuksissa voidaan säästää menettämättä liikaa tarkkuutta.
Tämä tutkielma käsittelee mukautuvaa hilantarkennusta menetelmänä optimoida 
simulointia Vlasiatorissa.

Tutkielmassa perehdytään plasman käyttäytymiseen ja erilaisiin ominaisiin mittakaavoihin, jotka tulee ottaa huomioon mallinnuksessa.
Mallinnusmenetelmistä käsitellään kineettistä mallinnusta tilastollisin menetelmin sekä fluidimallia.
Molemmissa on etunsa ja käyttökohteensa, ja Vlasiator käyttääkin näiden yhdistelmää.
Elektronien kineettisen mallinnus vaatii kertaluokkia tarkempaa resoluutiota kuin ioneiden, joten ionit mallinnetaan kineettisesti ja elektronit fluidina.

Vlasiatorin käyttämä kohdennettu tarkennus esitellään menetelmänä muisti- ja laskentavaatimuksissa säästämiseen.
Magnetosfäärin rakenteen vuoksi vaadittu tarkkuus ei ole sama kaikkialla si\-mu\-laa\-ti\-o\-a\-lu\-eel\-la.
Alueilla joissa vaaditaan parempaa resoluutiota simulaatiohilaa voidaan tarkentaa paikallisesti siten, että simulaatioon syötetään parametrinä tarkennettava alue.
Kehityksenä tästä esitellään mukautuva hilantarkennus -- menetelmä, millä simulaatio mukauttaa hilaa simulaatiodatan perusteella.
Mukautuva tarkennus perustuu eri muuttujien gradientteihin: alueet, joissa tarkasteltavat muuttujat muuttuvat äkillisesti, tarkennetaan.
Vastaavasti alueet, joissa muutos on vähäistä, harvennetaan.

Mukautuvan tarkennuksen tulokset osoittautuivat lupaavaksi tavaksi kehittää Vlasiatoria. 
Tarkennusparametrit tuottavat vanhaa tarkennusmenetelmää muistuttavan tuloksen, joka kuitenkin eroaa tarkalta muodoltaan siitä.
Seuraava jatkotutkimuksen kohde on mukautuvan tarkennuksen toteuttaminen ajonaikaisesti.

\end{abstract}

\mytableofcontents

\cleardoublepage
\pagenumbering{arabic}

\chapter{Johdanto}
Plasmaksi kutsutaan kvasineutraalia eli ulospäin varauksetonta, ionisoitunutta kaasua, jonka käyttäytymistä dominoivat kollektiiviset sähkömagneettiset vuorovaikutukset törmäysten sijasta. 
Enimmäkseen protoneista ja elektroneista koostuva aurinkotuuli on merkittävimpiä esimerkkejä avaruusplasmasta.
Plasman ja sähkömagneettisen kentän ilmiöitä kutsutaan avaruussääksi ja sen mallintaminen on tärkeää paitsi lähiavaruuden ymmärryksen vuoksi myös käytännön vaikutusten vuoksi.

Aurinkotuuli vaikuttaa elektroniikkaan ja kommunikaatioon sekä avaruustoiminnassa että pienemmissä määrin maan päällä.
Vuonna 1859 voimakkain mitattu aurinkomyrsky Carrington kaatoi lennätinverkon. 
Lähihistoriassa pienemmät aurinkomyrskyt ovat myös kaataneet paikallisia sähköverkkoja.
Uusi Carringtonin kaltainen myrsky voisi aiheuttaa mittavia ja pitkäkestoisia sähkökatkoja, katkaista mannertenväliset verkkoyhteydet tai tuhota satelliitteja \parencite{ylejuttu}.
Avaruussään luotettava ennustaminen on siis jatkuvassa määrin merkittävämpi tavoite.

Maan lähiavaruuden avaruustuulta mitataan pääasiassa satelliiteilla.
Näiden lähettäminen avaruuteen on kuitenkin kallista, ja mittauksilla on huono toistettavuus; mikäli satelliitti havaitsee tapahtuman, se on pian ohittanut sen eikä toista havaintoa välttämättä saada.
Ongelma voidaan ratkaista satelliittiparvilla, mutta tällöin kustannukset kasvavat \parencite{vlasiatorthesis}.

Lähimittausten kustannusten ja huonon toistettavuuden takia erilaiset simulaatiot ovat yksi tärkeimmistä menetelmistä aurinkotuulen mallinnukseen.
Keskimääräisen hiukkastiheyden ollessa luokkaa $\SI{E6}{\m\tothe{-3}}$ yksittäisten hiukkasten mallinnus on selkeästi epärealistinen lähestymistapa suuren mittakaavan simulaatioissa, joten plasmaa mallinnetaan yleensä karkeammalla tasolla:
hiukkasten sijaan voidaan käsitellä kuusiulotteisia paikka- ja nopeusjakaumia tai plasman keskimääräistä liikettä.
Vlasiator käyttää hybridisimulaationa näiden kahden menetelmän yhdistelmää.

Tutkielman rakenne on seuraavanlainen.
Kappaleessa \ref{ch:teoria} käsitellään plasman liikettä ja sen kuvausmenetelmiä. 
Kappaleessa \ref{ch:menetelmat} kuvaillaan Vlasiatoria, sen toimintaa, ja miten sitä voidaan parantaa mukautuvalla tarkennuksella.
Kappaleissa \ref{ch:tulokset} ja \ref{ch:johtopaatokset} käsitellään tämän menetelmän tuloksia.




\chapter{Teoriaa} \label{ch:teoria}
Kappaleet \ref{sec:plasmaliike} ja \ref{sec:skaalat} seuraavat kirjaa \citetitle{ipp} \parencite{ipp}. 
\section{Plasman liike} \label{sec:plasmaliike}
Varattuun hiukkaseen sähkömagneettisessa kentässä kohdistuu Lorentzin voima:
\begin{equation}
    \vb{F} = q \qty(\vb{E} + \vb{v} \cp \vb{B}), \label{eq:lorentz}
\end{equation}
missä $\vb{F}$ on voima, $q$ hiukkasen varaus, $\vb{E}$ sähkökenttä, $\vb{v}$ hiukkasen nopeus ja $\vb{B}$ magneettivuon tiheys.
Lisäksi plasma vaikuttaa sähkömagneettiseen kenttään Maxwellin yhtälöiden kautta:
\begin{align}
    \div{\vb{E}} &= \frac{\rho_\mathrm{c}}{\epsilon_0}                             \label{eq:gauss}    \\
    \div{\vb{B}} &= 0                                                   \label{eq:gauss2}   \\
    \curl{\vb{E}} &= -\pdv{\vb{B}}{t}                                   \label{eq:faraday}  \\
    \curl{\vb{B}} &= \mu_0 \qty(\vb{J} + \epsilon_0 \pdv{\vb{E}}{t}),    \label{eq:ampere}
\end{align}
missä $\rho_\mathrm{c}$ on varaustiheys, $\epsilon_0$ tyhjiön permittiivisyys, $t$ aika, $\mu_0$ tyhjiön permeabiliteetti ja $\vb{J}$ virrantiheys.
Kentät siis vaikuttavat plasman liikkeeseen ja plasman liike kenttiin.
Kenttien yhteydessä energiatiheys $U$ on myös tärkeä suure:
\begin{equation}
    U = \frac{\epsilon_0}{2} \vb{E}^2 + \frac{1}{2\mu_0} \vb{B}^2. \label{eq:energiatiheys0}
\end{equation}
Energiatiheyden yksikkö on \si{\J\per\m\cubed}.

Magnetosfäärissä myös taivaankappaleiden magneettiset dynamot vaikuttavat plasmaan:
konvektio ja pyörimisliike aiheuttavat esimerkiksi maan sulassa ytimessä virtoja, jotka luovat vahvan magneettikentän. 
Ampèren lain \eqref{eq:ampere} kautta plasman liikkeen synnyttämää magneettikenttää kutsutaan tässä yhteydessä perturboituneeksi magneettikentäksi $\vb{B}_1$ erotuksena taivaankappaleiden luomasta taustakentästä $\vb{B}_0$. 
Energiatiheydestä \eqref{eq:energiatiheys0} voidaan vähentää taustakentän kontribuutio ja tarkastella pelkästään perturboitunutta kenttää:
\begin{equation}
    U_1 = \frac{\epsilon_0}{2} \vb{E}^2 + \frac{1}{2\mu_0} \vb{B}_1^2. \label{eq:energiatiheys}
\end{equation}

\section{Ominaiset skaalat} \label{sec:skaalat}
\subsection{Langmuirin värähtely}
Poikkeama elektronitiheydessä aiheuttaa myös liikettä plasmassa. 
Olettamalla raskaammat ionit liikkumattomaksi elektronien taustalla, saadaan häiriölle $n_1(\vb{r}, t)$ ja hiukkastiheydelle $n_0$:
\begin{align*}
    \begin{cases}
        n_\mathrm{i} = n_0  \\
        n_\mathrm{e} = n_0 + n_1(\vb{r}, t),
    \end{cases}
\end{align*}
missä $n_\mathrm{i}$ on ionien ja $n_\mathrm{e}$ elektronien lukumäärätiheys.
Tämä perturbaatio aiheuttaa sähkökentän $\vb{E}_1$.
Elektronien liikeyhtälö on nyt:
\begin{align}
    \vb{F} &= m \vb{a}  && \left|\text{yhtälöstä } \eqref{eq:lorentz}\right.   \notag \\
    q_\mathrm{e} \vb{E}_1 &= m_\mathrm{e} \pdv{\vb{u}_1}{t}, \label{eq:vittu0}
\end{align}
missä $\vb{a}$ on kiihtyvyys, $m_\mathrm{e}$ elektronin massa, $q_\mathrm{e}$ elektronin varaus eli alkeisvaraus $e$ ja $\vb{u}_1$ perturbaation aiheuttama liike.
Jatkuvuusyhtälöstä saadaan olettamalla perturbaation olevan pieni ja poistamalla toisen asteen termi $n_1 \vb{u}_1$:
\begin{align}
    0 &= \pdv{n_\mathrm{e}}{t} + \div(n_\mathrm{e} \vb{u})    \notag \\
    0 &= \pdv{n_1}{t} + n_0 \div\vb{u}_1  
        && \left|\pdv{t}\right.  \notag   \\
    0 &= \pdv[2]{n_1}{t} + n_0 \div(\pdv{\vb{u}_1}{t})
        && \left|\text{yhtälöstä } \eqref{eq:vittu0}\right. \notag \\
    0 &= \pdv[2]{n_1}{t} + \frac{n_0 q_\mathrm{e}}{m_\mathrm{e}} \div\vb{E}_1
        && \left|\text{yhtälöstä } \eqref{eq:gauss}\right.   \notag \\
    0 &= \pdv[2]{n_1}{t} + \frac{n_0 q_\mathrm{e} \rho_\mathrm{c, 1}}{\epsilon_0 m_\mathrm{e}}
        && \left|\rho_\mathrm{c, 1} = q_\mathrm{e} n_1 \right. \notag \\
    0 &= \pdv[2]{n_1}{t} + \frac{n_0 q_\mathrm{e}^2}{\epsilon_0 m_\mathrm{e}} n_1
        && \left|q_\mathrm{e} = e\right.  \notag \\
    0 &= \pdv[2]{n_1}{t} + \frac{n_0 e^2}{\epsilon_0 m_\mathrm{e}} n_1. \label{eq:liikeyhtalo0}
\end{align}
Yhtälöstä \label{eq:liikeyhtalo0} nähdään liikkeen olevan harmonista värähtelyä taajuudella
\begin{equation}
    \omega_\mathrm{pe}^2 = \frac{n_0 e^2}{\epsilon_0 m_\mathrm{e}}, \label{eq:plasmataajuus}
\end{equation}
jota kutsutaan \emph{plasmataajuudeksi}. 
Vastaavasti ioneille voidaan määrittää hitaampi värähtely:
\begin{align}
    \omega_\mathrm{pi}^2 &= \frac{n_0 q_\mathrm{i}^2}{\epsilon_0 m_\mathrm{i}}  \notag  \\
    \omega_\mathrm{pi}^2 &= \frac{n_0 \qty(Z_\mathrm{i} e)^2}{\epsilon_0 m_\mathrm{i}}. \label{eq:ioniplasmataajuus}
\end{align}
missä $q_\mathrm{i}$ on ionin varaus, $Z_\mathrm{i}$ ionin järjestysluku ja $m_\mathrm{i}$ ionin massa.
Plasmataajuudet muodostavat plasman ominaisen aikaskaalan.

\subsection{Larmorin liike}
Tarkastellaan nyt liikkuvaa hiukkasta sähkökentässä $\vb{E} = \vb{0}$ ja magneettikentässä $\vb{B} = B \vu{k}$.
Hiukkasen massa on $m$ ja varaus $q$.
Tällöin hiukkaseen kiihtyvyys kuhunkin suuntaan saadaan:
\begin{align*}
    m \vb{a} &= \vb{F}  && \left|\text{Yhtälöstä } \eqref{eq:lorentz} \right. \\
    \pdv{\vb{v}}{t} &= \frac{q}{m} \vb{B} \cp \vb{v}  \\
    \pdv{\vb{v}}{t} &= \frac{qB}{m} \qty(v_y, -v_x, 0)    
        && \left|\pdv{t}\right.\\
    \pdv[2]{\vb{v}}{t} &= \frac{qB}{m} \qty(\pdv{v_y}{t}, -\pdv{v_x}{t}, 0) \\
    \pdv[2]{\vb{v}}{t} &= -\frac{q^2 B^2}{m^2} \qty(v_x, v_y, 0)
\end{align*}
Hiukkasen kentän suuntainen nopeus on siis vakio, ja kenttää kohtisuoraan se päätyy ympyräradalle taajuudella
\begin{equation}
    \omega_\mathrm{c} = \frac{q B}{m}, \label{eq:gyrofreq}
\end{equation}
jota kutsutaan \emph{Larmorin taajuudeksi} eli \emph{gyrotaajuudeksi}. 
Ympyräradan säde eli \emph{Larmorin säde} $r_\mathrm{L}$ ratkaistaan melko suoraviivaisesti:
\begin{align}
    \vb{F} &= m \vb{a}  && \left|\text{yhtälöstä } \eqref{eq:lorentz} \text{ ja normaalikiihtyvyydestä}\right. \notag \\
    q v_\perp B &= m \frac{v_\perp^2}{r_\mathrm{L}}  \notag \\
    r_\mathrm{L} &= \frac{m v_\perp}{q B}, \label{eq:gyroradius}
\end{align}
missä $v_\perp$ on hiukkasen nopeuden suuruus magneettikenttää kohtisuoraan.

Voimme myös määrittää \emph{inertiaalipituuden} Larmorin säteen ja niin kutsutun \emph{Alfvénin nopeuden} avulla. 
Alfvénin nopeus $v_\mathrm{A}$ on magneettikentän suuntaan kulkevan oskillaation nopeus, ja se määritellään
\[
    v_\mathrm{A} = \frac{B}{\sqrt{\mu_0 \rho}}.
\]
Inertiaalipituus $d$ on Alfvénin nopeudella kulkevan hiukkasen Larmorin säde:
\begin{align}
    d &= \frac{m B}{q B \sqrt{\mu_0 \rho}}    \notag \\
      &= \frac{m c \sqrt{\epsilon_0}}{q \sqrt{n_0 m}}    \notag \\
      &= \frac{c}{q} \sqrt{\frac{m \epsilon_0}{n_0}}
        && \left|\omega_\mathrm{p}\right.    \notag \\
      &= \frac{c}{\omega_\mathrm{p}}, \label{eq:inertiaalipituus}
\end{align}
missä $c$ on valon nopeus tyhjiössä ja $\omega_\mathrm{p}$ kyseisen hiukkasen plasmataajuus \eqref{eq:plasmataajuus} tai \eqref{eq:ioniplasmataajuus}.

Inertiaalipituus ja Larmorin säde muodostavat kineettisten ilmiöiden etäisyysskaalan:
näitä lyhyemmillä pituuksilla kineettiset ilmiöt eli yksittäisten hiukkasten välisiin vuorovaikutukset tulevat esiin \parencite{vlasiatorthesis}.
Samoin käy plasmataajuutta sekä gyrotaajuutta lyhyemmillä aika-askeleilla.

Larmorin liikkeen vuoksi plasma myös virtaa anisotrooppisesti:
hiukkaset liikkuvat vapaasti magneettikentän suuntaisesti, mutta magneettikenttää vastaan kohtisuora liike on vaikeaa.

\section{Aurinkotuulen ilmiöt}
Plasma voidaan jakaa törmäykselliseen ja törmäyksettömään plasmaan. 
Aurinkotuulta voidaan kuvata törmäyksettömänä; hiukkasten keskimääräinen vapaa matka on \SI{1}{AU} kokoluokkaa \parencite{ipp}.



Maan magneettikenttä estää aurinkotuulen suoraa virtausta, jolloin hiukkaset joutuvat kiertämään magnetosfäärin.
Aurinkotuulen ionien virtausnopeus on noin \SI{400}{\km\per\s}, mikä on enemmän kuin perturbaatioiden etenemisnopeus plasmassa.
Auringon puolelle Maata eli päiväpuolelle muodostuu tällöin keulasokki jossa nopeus äkillisesti laskee -- ilmiö on vastaava kuin yliäänikoneilla ilmassa \parencite{shocks}.

Maan magneetosfäärin ja aurinkotuulen välistä rajapintaa kutsutaan magnetopausiksi.
Aurinkotuulen virtauspaine luo magnetosfäärille ominaisen muodon:
Päiväpuolella magnetosfääri on litistynyt, ja Auringon vastaisella puolella eli yöpuolella magnetosfääri venyy pitkäksi pyrstöksi.
Magnetopausin ja keulasokin välissä on huotra, jossa aurinkotuuli kiertää magnetosfäärin \parencite{isp}.

Keulasokin vuoksi aurinkotuuli käyttäytyy eri tavoilla eri alueilla. 
Kaukana magnetosfääristä virta on hyvin tasaista, kun taas Maan vaikutuspiirissä aurinkotuulen tiheys tai nopeus voivat muuttua hyvin äkillisesti.

\begin{figure}[p]
    \centering
    \includegraphics[width=\textwidth]{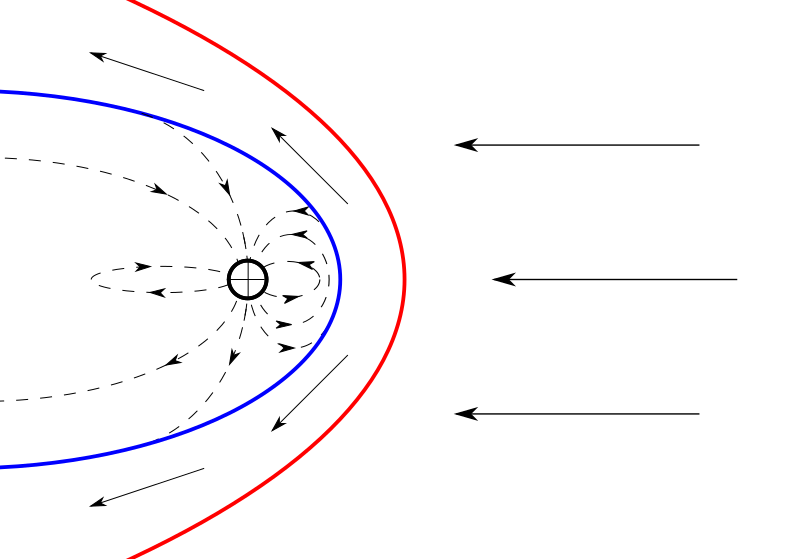}
    \caption{
        Yksinkertaistettu kuva Maan ($\oplus$) magnetosfääristä, mukaillen kirjaa \protect\citetitle{isp} \protect\parencite[Kuva 9.3]{isp}.  
        Sininen viiva on magnetopausi; Maan magnetosfääri on sen sisäpuolella.
        Magnetopausin ja punaisen keulasokin välissä on huotra.
        Kiinteät nuolet kuvaavat aurinkotuulen virtaussuuntaa ja katkoviivat Maan magneettikenttää.
    }
    \label{fig:unrefined}
\end{figure}


\chapter{Menetelmät} \label{ch:menetelmat}
\section{Plasman kuvaus}
\subsection{Tilastollinen kuvaus}
Tilastollisessa kuvauksessa käytetään statistisen mekaniikan menetelmiä plasman kineettiseen kuvaamiseen.
Yksittäisten hiukkasten käyttäytymisen mallintamisen sijaan käsitellään kuusiulotteista distribuutiofunktiota $f(\vb{x}, \vb{v}, t)$, jolle:
\begin{align*}
    \iiint \dd[3]{\vb{v}} f(\vb{r}, \vb{v}) &= n(\vb{x}, t)    \\
    \iiint \dd[3]{\vb{x}} \iiint \dd[3]{\vb{v}} f(\vb{x}, \vb{v}, t) &= N(t),
\end{align*}
missä $\vb{x}$ on paikka, $\vb{v}$ nopeus, $n$ hiukkastiheys ja $N$ hiukkasten kokonaislukumäärä.

Tarkastellaan nyt törmäyksettömässä tapauksessa tilavuuselementtiä $\mathcal{V}$, jolle:
\begin{align}
    N_\mathcal{V}(t) &= \int_\mathcal{V} \dd[3]{\vb{v}} \dd[3]{\vb{x}} f(t, \vb{x}, \vb{v})     \notag\\
    \dv{t} N_\mathcal{V}(t) &= \int_\mathcal{V} \dd[3]{\vb{v}} \dd[3]{\vb{x}} \pdv{t} f(t, \vb{x}, \vb{v}). \label{eq:paska1}
\end{align}
Koska hiukkasia ei tuhoudu tai muodostu, hiukkasmäärän muutos tilavuuselementissä johtuu hiukkasvuosta pinnan $\mathcal{S} = \partial \mathcal{V}$ läpi.
Merkitään nyt $\vb*{\xi} = (\vb{x}, \vb{v})$ ja $\vb{u} = \dot{\vb*{\xi}}$. Nyt saadaan:
\begin{align}
    \dv{t} N_\mathcal{V} &= -\oint_\mathcal{S} \vb{u} f \vdot \dd{\vb{\mathcal{S}}} 
        && \left|\text{Divergenssiteoreemalla:}\right. \notag\\
    \dv{t} N_\mathcal{V} &= - \int_\mathcal{V} \grad_{\vb*{\xi}} \vdot (\vb{u} f) \dd{\mathcal{V}}. \label{eq:paska2}
\end{align}
Vähentämällä yhtälöstä \eqref{eq:paska1} yhtälö \eqref{eq:paska2} saadaan:
\[
    \int_\mathcal{V} \qty[\pdv{f}{t} + \grad_{\vb*{\xi}} \vdot (\vb{u} f)] \dd{\mathcal{V}} = 0
\]
Tämä pätee kaikille $\mathcal{V}$, joten:
\begin{align}
    0 &= \pdv{f}{t} + \grad_{\vb*{\xi}} \vdot (\vb{u} f)  \notag \\
    0 &= \pdv{f}{t} + \grad_{\vb{x}} \vdot (\dot{\vb{x}} f) + \grad_{\vb{v}} \vdot (\dot{\vb{v}} f)  \notag \\
    0 &= \pdv{f}{t} + \grad_{\vb{x}} \vdot (\vb{v} f) + \grad_{\vb{v}} \vdot \qty(\frac{\vb{F}}{m} f)  \notag \\
    0 &= \pdv{f}{t} + \cancel{f \grad_{\vb{x}} \vb{v}} + \vb{v} \vdot \grad_{\vb{x}} f + f \grad_{\vb{v}} \vdot \frac{\vb{F}}{m} + \frac{\vb{F}}{m} \vdot \grad_{\vb{v}} f    \notag \\
    0 &= \pdv{f}{t} + \vb{v} \vdot \grad_{\vb{x}} f + f \grad_{\vb{v}} \vdot \frac{\vb{F}}{m} + \frac{\vb{F}}{m} \vdot \grad_{\vb{v}} f. \label{eq:generalvlasov}
\end{align}
Termi $\grad_{\vb{v}} \vdot \vb{F}$ on nolla mikäli voima ei riipu nopeudesta. 
Lorentzin voima riippuu, mutta sille:
\begin{align*}
    \grad_{\vb{v}} \vdot \vb{F} &= q \grad_{\vb{v}} \vdot (\vb{E} + \vb{v} \cp \vb{B})    \\
    \grad_{\vb{v}} \vdot \vb{F} &= q \sum_{i, j, k} \pdv{v_i} (\epsilon_{ijk} v_j B_k)  \\
    \grad_{\vb{v}} \vdot \vb{F} &= 0.
\end{align*}
Sijoittamalla nyt Lorentzin voima \eqref{eq:lorentz} yhtälöön \eqref{eq:generalvlasov} saadaan siis:
\begin{equation}
    \pdv{f}{t} + \vb{v} \vdot \grad_x{f} + \frac{q}{m} \qty(\vb{E} + \vb{v} \cp \vb{B}) \vdot \grad_{\vb{v}} f = 0.   \label{eq:vlasov}
\end{equation}
Tätä kutsutaan \emph{Vlasovin yhtälöksi}. Menetelmässä iteroidaan Vlasovin yhtälöä aika-askel kerrallaan. 
Kentät määritetään myös aika-askeleittain Maxwellin yhtälöiden avulla.

\subsection{Magnetohydrodynaminen kuvaus}
Fluidikuvaus on kineettisiä kuvauksia yksinkertaisempi.
Määrittämällä jakaumafunktion nopeusmomentit voidaan plasman liikettä kuvailla esimerkiksi hiukkastiheyden, virtausnopeuden ja paineen avulla.
Käytännössä hiukkaslajit yksinkertaistetaan homogeenisiksi fluideiksi ja liikeyhtälöt muistuttavat Navier-Stokes yhtälöiden ja Maxwellin yhtälöiden yhdistelmää \parencite{ipp}.

Tilastolliseen kuvaukseen verrattuna nopeusavaruus yksinkertaistuu kokonaan pois.
Menetelmä on täten merkittävästi kevyempi, mutta malli antaa vain pääpiirteisen kuvan plasman liikkeestä.
Esimerkiksi plasman läpi virtaava, ympäröivästä jakaumasta poikkeava suihku aiheuttaa epästabiiliuksia joita fluidimalli ei kykene kuvaamaan \parencite{vlasiatorthesis}.
Kineettisissä kuvauksissa on myös aaltomoodeja, jotka eivät ilmene fluidikuvauksessa \parencite{wavedispersion}.

Yhdistämällä kaikki hiukkastyypit yhdeksi fluidiksi päästään \emph{magnetohydrodynaamiseen} kuvaukseen eli MHD:hen.
Tässä kuvauksessa myös hiukkastyyppien väliset vuorovaikutukset katoavat; 
elektronit ja kaikki ionit virtaavat samansuuntaisesti samalla nopeudella, mikä edelleen yksinkertaistaa kuvausta.
Fluidimallit toimivat parhaiten termodynaamisessa tasapainossa olevaan plasmaan, mutta menetelmää on käytetty myös aurinkotuulen kuvaukseen.
Esimerkkinä tästä on GUMICS-simulaatio \parencite{gumics}.

\section{Vlasiator}
Vlasiator\footnote{Tuoreimman version lähdekoodi saatavilla \parencite{vlasiatorcode}} käyttää magnetosfäärin globaaliin mallinnukseen niin sanottua hybridi-Vlasov-menetelmää.
Ionit mallinnetaan kineettisellä kuvauksella ja elektronit fluidina \parencite{vlasiatorpaper}.
Menetelmän hyöty juontuu elektronien pienemmästä massasta. 
Kuten kappaleessa \ref{ch:teoria} on johdettu, Larmorin liikkeen taajuus \eqref{eq:gyrofreq} ja säde \eqref{eq:gyroradius} sekä plasmataajuus \eqref{eq:plasmataajuus} ja inertiaalipituus \eqref{eq:inertiaalipituus} riippuvat massasta seuraavalla tavalla:
\[
    \frac{\omega_\text{ce}}{\omega_{ci}} = \frac{\omega_\text{pe}^2}{\omega_\text{pi}^2} = \frac{r_\text{Li}}{r_\text{Le}} = \frac{d_i^2}{d_e^2} = \frac{m_\mathrm{i}}{m_\mathrm{e}} \approx \num{1836}.
\]
Vetyionin massa on noin \num{1836} suurempi kuin elektronin, joten elektronien kineettinen mallintaminen vaatii paljon lyhyemmän aika-askeleen sekä tarkemman resoluution ja täten kohtuuttoman paljon laskenta-aikaa \parencite{vlasiatorthesis}.
Hybridimenetelmällä yhdistetään kineettiset ilmiöt ioneilla fluidikuvauksen tehokkuuteen.

\section{Simulaation vaatimukset}
Globaalien ilmiöiden tarkkaa mallinnusta varten simulaation fyysinen koko on varsin suuri. 
Otetaan esimerkiksi kuun kiertosäteen verran eli noin \num{60} Maan sädettä tilaa jokaiseen suuntaan ($x, y, z \in [-60 R_\mathrm{E}, 60 R_\mathrm{E}]$).
Kineettisten ilmiöiden mallintamiseen resoluution tulee olla samaa kokoluokkaa kuin ionien inertiaalinen pituus aurinkotuulessa $d_i \approx \SI{100}{\km}$ \parencite{resolution}.
Paikka-avaruuden soluja $N_x$ tarvitaan siis yhteensä:
\[
    N_x = \qty(\frac{2 \cdot 60 \cdot \SI{6371}{\km}}{\SI{100}{\km}})^3 \approx \num{4E11}
\]
Nopeusavaruuden tulee taas kattaa aurinkotuulelle tyypilliset nopeudet, eli joka suuntaan $\SI{\pm 2000}{\km\per\s}$.
Resoluution tulee olla aurinkotuulen termisen nopeuden luokkaa eli noin \SI{30}{\km\per\s} \parencite{resolution}.
Nopeusavaruuden soluja $N_v$ tarvitaan siis:
\[
    N_v = \qty(\frac{\SI{2000}{\km\per\s}}{\SI{30}{\km\per\s}})^3 \approx \num{3E5}
\]
Yhteensä faasiavaruus on siis $3 \cdot \num{4E16} \approx \num{1E17}$ solua, joka on nykyisille super\-tieto\-koneille liikaa.

Merkittävä osuus faasiavaruudesta on kuitenkin erittäin harvaa eikä täysi tarkkuus ole täten tarpeellista kineettisten ilmiöiden mallintamiseen. 
Harventamalla nopeusavaruutta päästään \SI{98}{\percent} säästöihin muisti- ja laskentavaatimuksissa \parencite{vlasovmethods}. 
Tämä riittää kaksiulotteisiin simulaatioihin, mutta kolmiulotteisessa tapauksessa vaadittuun tarkkuuteen tarvitaan myös paikka-avaruuden harventamista.

\section{Kohdennettu tarkennus}
Kaukana Maasta plasma on varsin homogeenista verrattuna Maan magnetosfäärin alueeseen, ks. Kuva \ref{fig:unrefined}.
Täten kaikkien solujen pitäminen samalla tarkkuudella on tarpeetonta; ylimääräinen tarkkuus tasaisella alueella kuluttaa resursseja antamatta hyödyllistä informaatiota simulaatiota varten.
Koko simulaation nostaminen esimerkiksi kaksi kertaa tarkemmaksi johtaa $2^3 = 8$ -kertaiseen muistinkäyttöön ja laskentamäärään.

\begin{figure}[p]
    \centering
    \includegraphics[width=0.9\textwidth]{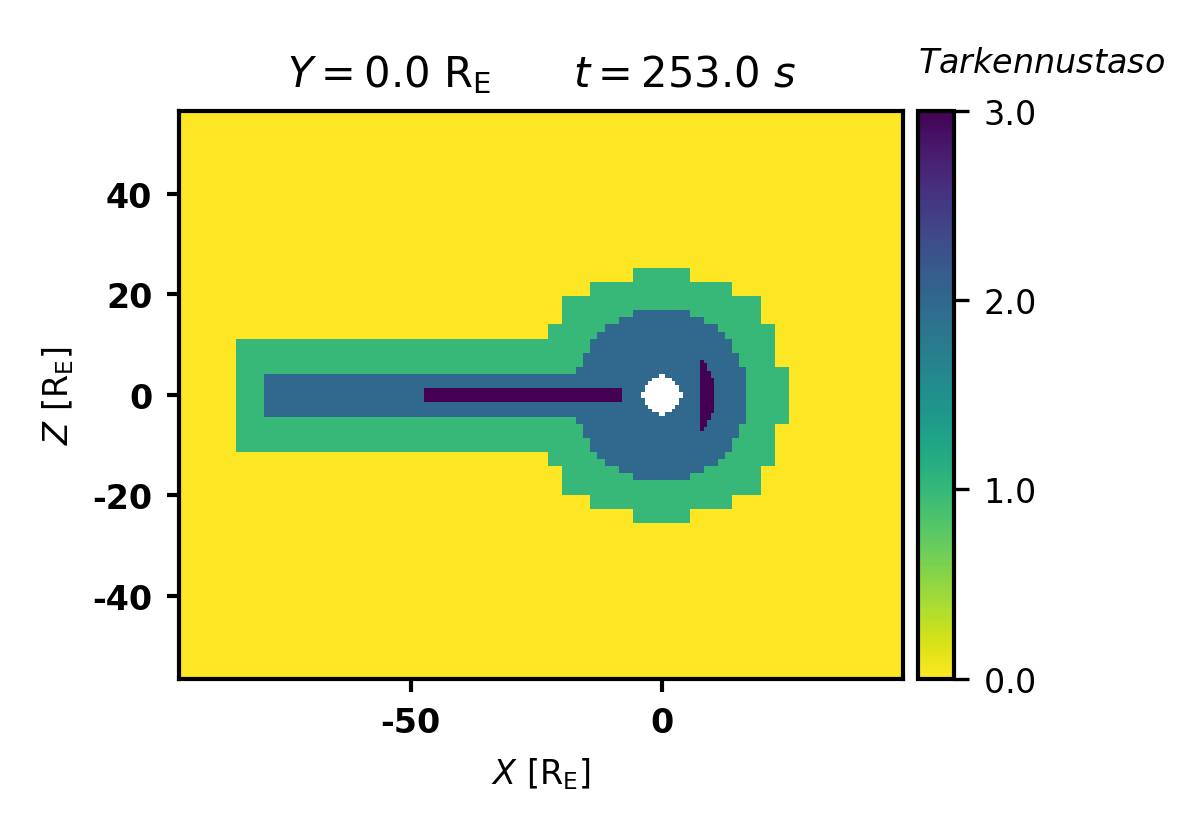}
    \caption{Esimerkki parametrisoidusta tarkennuksesta, tummempi väri tarkoittaa tarkempaa resoluutiota. Häntä ja magnetosfäärin kärki ovat tarkimmalla tasolla.}
    \label{fig:oldrefinement}
\end{figure}

Vlasiatorissa ongelma on ratkaistu dccrg-kirjastolla.\footnote{Distributed Cartesian cell refinable grid, lähdekoodi \url{https://github.com/fmihpc/dccrg}}
Kukin solu voidaan kolmessa ulottuvuudessa jakaa kahdeksaan pienempään soluun.
Kahden vierekkäisen solun tarkennukset voivat erota enintään yhdellä tasolla \parencite{dccrgpaper}.
Toistaiseksi simulaatiota ajaessa on annettu parametreina tarkennettava alue:
ionosfäärin ympäristö ja magnetosfäärin pyrstö ovat tarkimmalla tasolla, ks. Kuva \ref{fig:oldrefinement}. 


Tämä on toimiva joskaan ei täysin ongelmaton ratkaisu. 
Ensinnäkin vaikka parametrit vastaavatkin Maan magnetosfäärin rakenteita, ne ovat jokseenkin mielivaltaisesti määriteltyjä.
Etukäteen määritellyillä parametreilla tarkennettava alue saattaa olla tarpeettoman suuri, tuhlaten laskentaresursseja.
Toisaalta sen ulkopuolelle voi myös jäädä turbulentteja alueita, joissa tarkempi kuvaus olisi hyödyllistä.
Parametrisaatio ei myöskään ole kovin yleistettävä.
Parametrit tulisi määrittää erikseen esimerkiksi muille taivaankappaleille tai erilaisille simulaatio-olosuhteille.

\begin{figure}[p]
    \centering
    \includegraphics[width=0.9\textwidth]{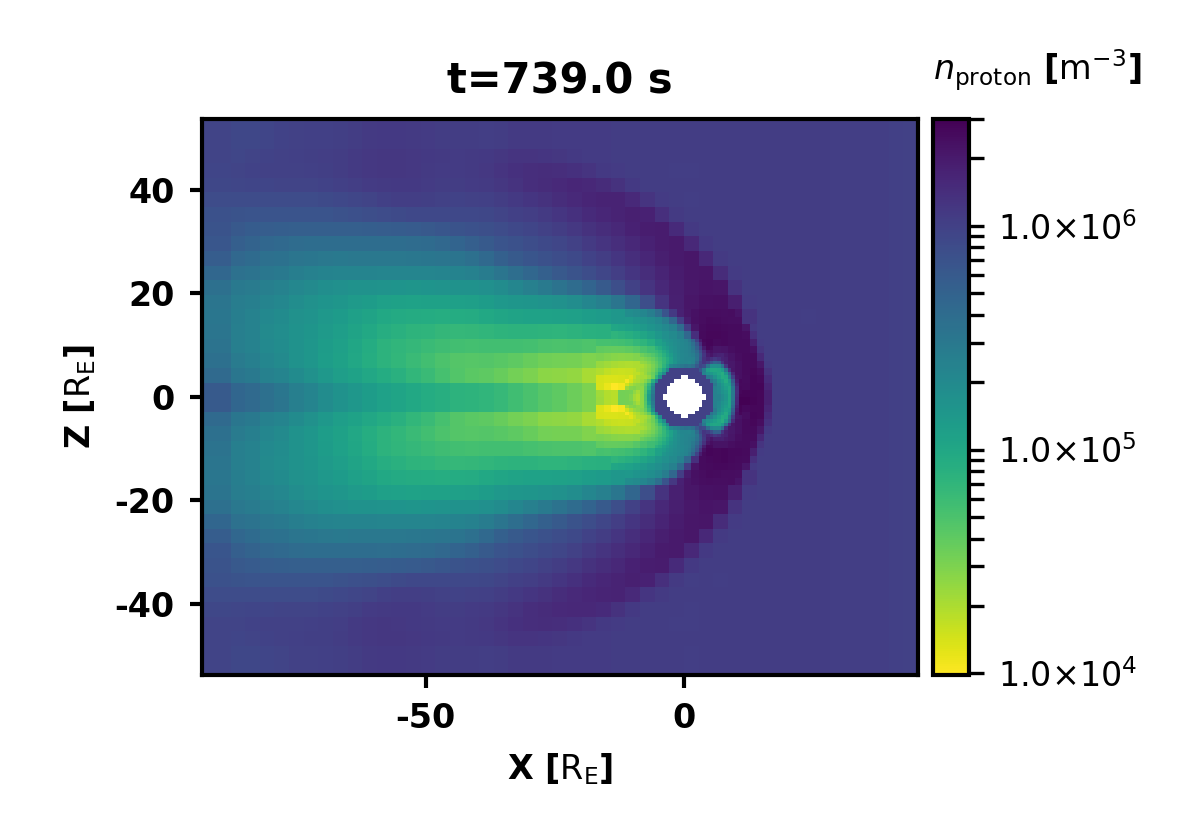}
    \caption{Esimerkki protonien hiukkastiheyden $n$ käyttäytymisestä, kun ainoastaan ionosfäärin ympäristö on korkealla resoluutiolla.}
    \label{fig:unrefined}
\end{figure}

\pagebreak
\section{Mukautuva hilatarkennus}
Ratkaisuna tähän on mukautuva hilantarkennus eli AMR\footnote{Adaptive mesh refinement}.
Simulaation aikana voidaan määrittää jokaisessa solussa niin sanottu tarkennusindeksi $\alpha$, jonka perusteella hilaa joko tihennetään tai harvennetaan.
Nyt simulaatiolle annetaan parametrina kaksi kynnystä. Indeksin alittaessa alemman kynnyksen koordinaatistoa harvennetaan solun kohdalta ja sen ylittäessä ylemmän kynnyksen sitä tihennetään.

Tarkennusindeksi on GUMICS-simulaatiossa määritetty dimensiottomien gradienttien maksimina seuraavalla tavalla \parencite{gumics}:
\begin{equation}
    \alpha = \mathrm{max} \qty{\frac{\Delta \rho}{\hat{\rho}}, \frac{\Delta U_1}{\widehat{U_1}}, \frac{\qty(\Delta \vb{p})^2}{2\rho \widehat{U_1}}, \frac{\qty(\Delta \vb{B}_1)^2}{2\mu_0 \widehat{U_1}}, \frac{\abs{\Delta \vb{B}_1}}{\widehat{B_1}}}. \label{eq:alpha}
\end{equation} 
$\Delta x$ tässä on kahden solun välinen erotus muuttujassa $x$ ja $\hat{x}$ maksimi niiden välillä. 
Tarkennusindeksiä määrittäessä lasketaan jokainen termi tarkastelemalla solua ja kutakin sen naapureista yksitellen.
Tarkasteltavina muuttujina ovat tiheys $\rho$, perturboituneen kentän energiatiheys $U_1$ \eqref{eq:energiatiheys}, liikemäärätiheys $\vb{p}$, ja perturboituneen magneettikentän tiheys $\vb{B}_1$.

Tämä tarkennusindeksi otettiin käyttöön myös Vlasiatorissa.
Tarkennuksessa solu jaetaan kahdeksaan tytärsoluun ja data yksinkertaisesti kopioidaan emosolulta niille. 
Harvennuksessa otetaan kahdeksasta solusta keskiarvo, joka asetetaan suuremman solun arvoksi.

Tämän lisäksi voidaan myös suodattaa data liukuvan keskiarvon avulla.
Tarkennettujen solujen datan arvoksi asetetaan keskiarvo niiden omasta ja tarkennettujen naapurien datasta.
Tämä tekee uudelleentarkennetusta datasta hieman sileämpää, säilyttäen kuitenkin esimerkiksi ionien kokonaisvarauksen ja -liikemäärän.

\chapter{Tulokset} \label{ch:tulokset}
Tarkennus toteutettiin indeksin $\alpha$ \eqref{eq:alpha} avulla pysäyttämällä ensin tasaisella resoluutiolla ajettu simulaatio (kuva \ref{fig:unrefined}).
Tämän jälkeen tarkennus tehdään käynnistäessä simulaatio uudelleen.
Useampi tarkennustaso toteutettiin jakamalla tarkennusindeksi kahdella aina tarkentaessa, ja tarkentamalla uudestaan mikäli tarkennuskynnys edelleen ylittyi.
Kaikki ajoparametrit ovat Liitteessä \ref{ch:parameters}. 
Tarkentaessa kohdan \verb|[restart]| alle lisättiin edellisen tiedoston nimi ja muutettiin \verb|adapt_refinement = 1| rivillä 10.
Kahden tarkennustason tulos saadaan muuttamalla \verb|max_spatial_level = 2| rivillä 8.

\begin{figure}[p]
    \includegraphics[width=\textwidth]{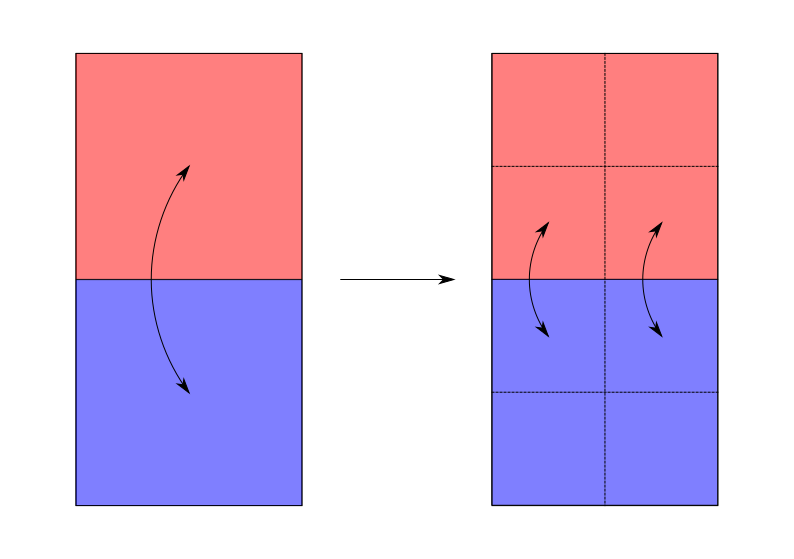}
    \caption{
        Havainnollistus tarkennusindeksin käyttäytymisestä välittömästi tarkennuksen jälkeen. 
        Olkoon punaisen (yllä) ja sinisen (alla) solun välillä sellainen ero, että tarkennusindeksi $\alpha$ ylittää tarkennuskynnyksen (vasemmalla).
        Tarkennuksessa data kopioituu suoraan tytärsoluille.
        Tarkennuksen jälkeen punaisen solun alempien tytärsolujen ja sinisen solun ylempien tytärsolujen välinen ero on sama kuin emosolujen välillä,
        joten tarkennusindeksi edelleen ylittyy (oikealla).
    }
    \label{fig:siistipiirros}
\end{figure}

Tarkennusta ei tehty ajon aikana useista syistä. 
Ajo oli ensisijaisesti testi tarkennusparametreille.
Tarkennuksen suorittaminen kenttien alustamisen yhteydessä vähensi mahdollisuutta virheille itse simulaatiossa.
Tarkennuksen aikavaatimusta ei vielä oltu määritetty ja testin tarkoituksena oli myös kartoittaa tältä osin soveltuvuutta ajonaikaiseen tarkennukseen.
Välittömästi tarkennuksen jälkeen tarkennusindeksi ei myöskään ole enää hyvä mittari, sillä data kopioituu suoraan tytärsoluille.
Kuvassa \ref{fig:siistipiirros} on havainnollistus ongelmasta.

\begin{figure}[p]
    \centering
    \includegraphics[width=0.9\textwidth]{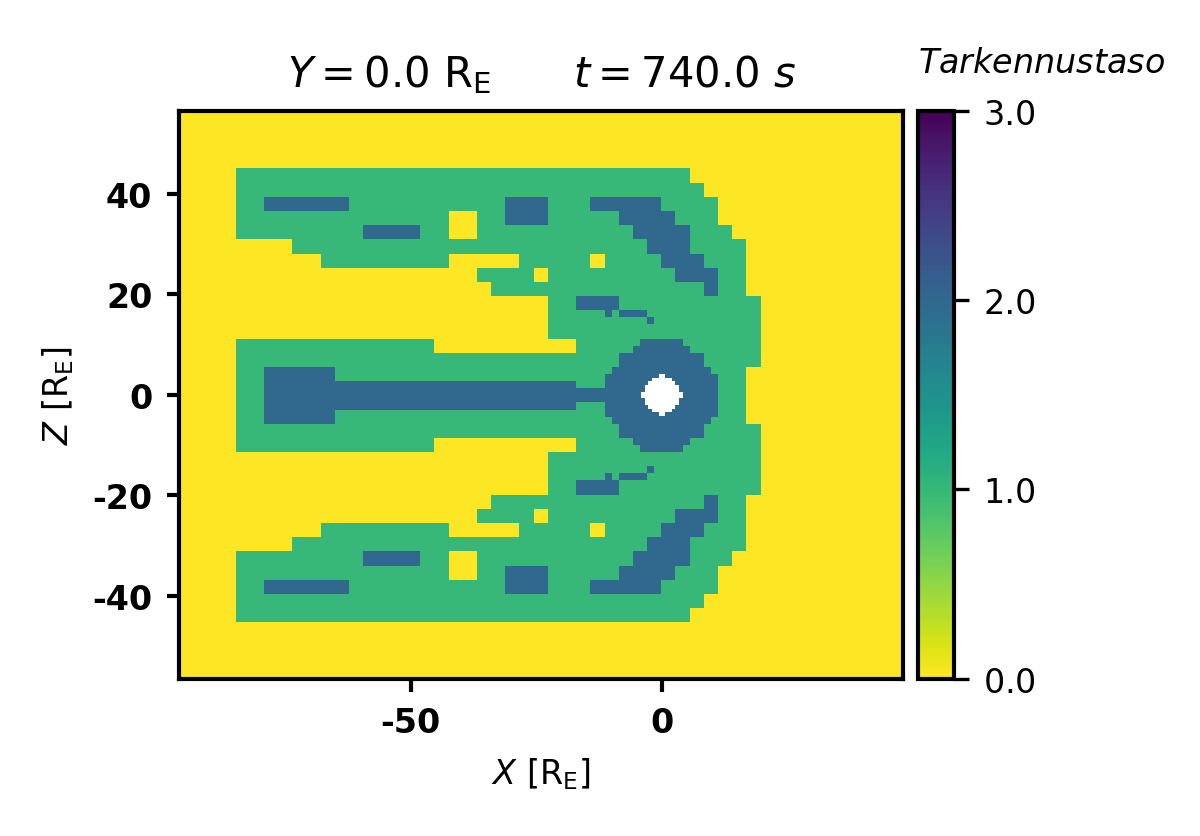}
    \caption{Mukautuvan tarkennuksen tulos kahdella tarkennustasolla.} 
    \label{fig:reflevel2}
\end{figure}

\begin{figure}[p]
    \centering
    \includegraphics[width=0.9\textwidth]{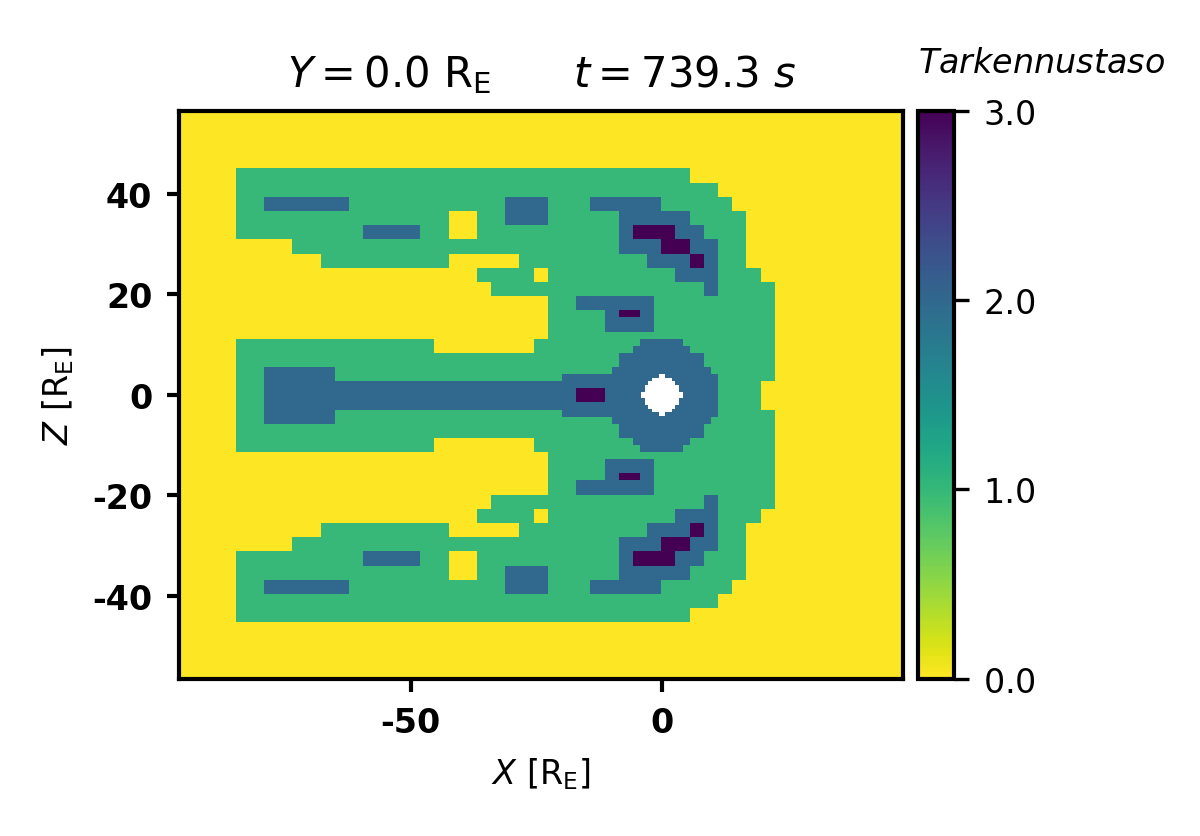}
    \caption{Mukautuvan tarkennuksen tulos kolmella tarkennustasolla.} 
    \label{fig:reflevel3}
\end{figure}

Kuvissa \ref{fig:reflevel2} ja \ref{fig:reflevel3} näkyy lopputulos: verrattuna parametrisoituun tarkennukseen Kuvassa \ref{fig:oldrefinement} mukautuva tarkennus keskittyy enemmän keulasokin sivuihin kärjen ja hännän sijasta.

\begin{figure}[p]
    \centering
    \includegraphics[width=0.9\textwidth]{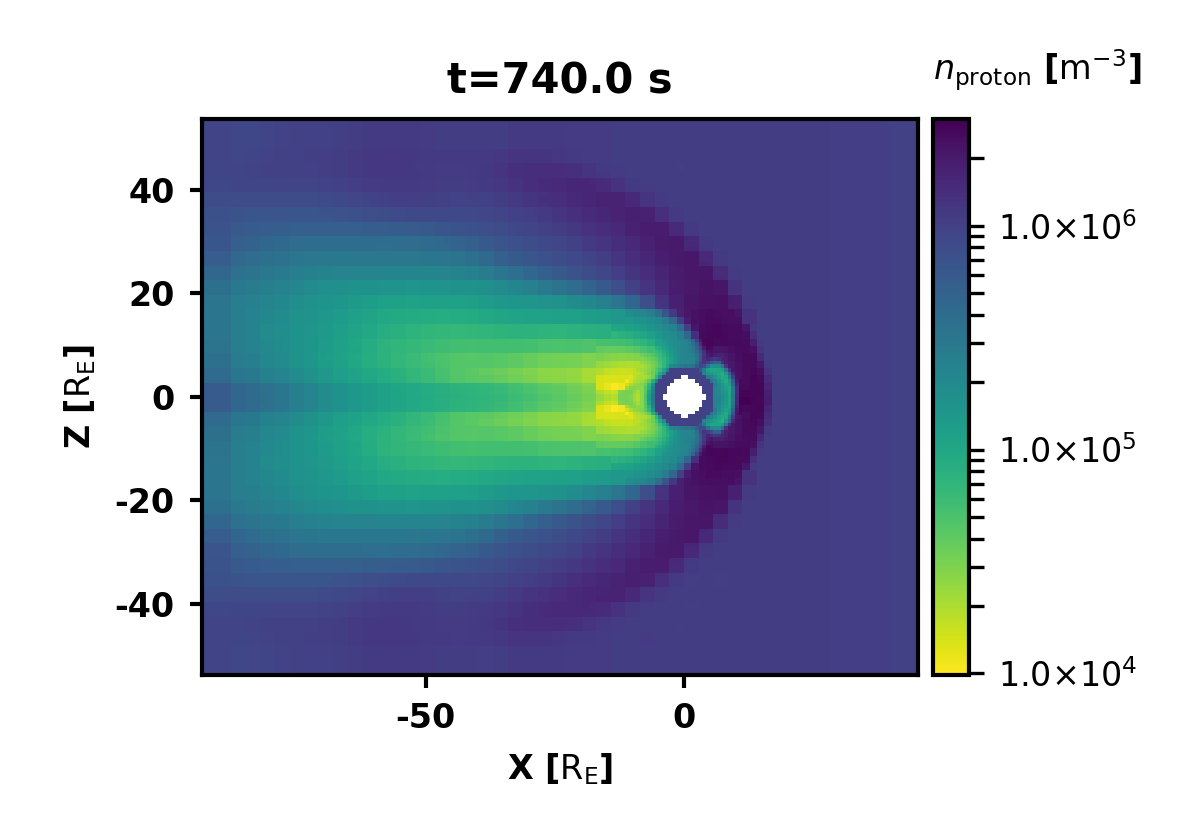}
    \includegraphics[width=0.9\textwidth]{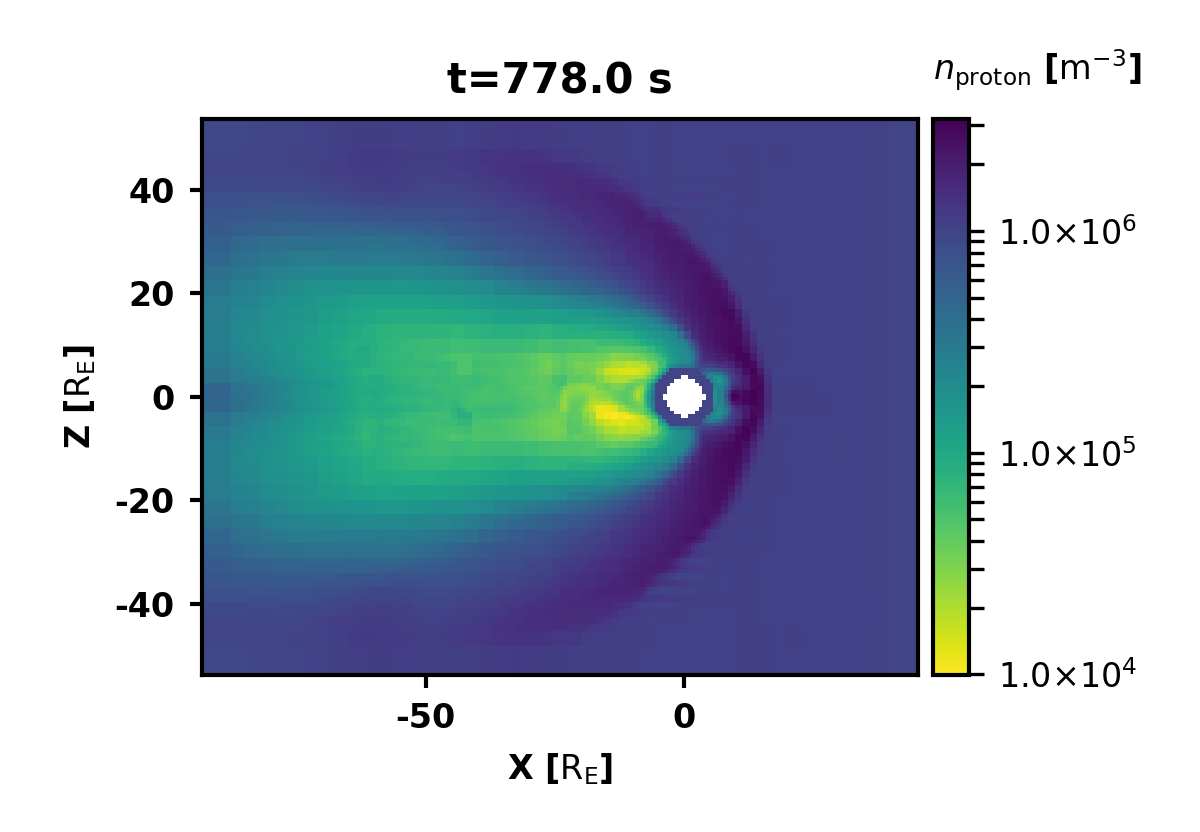}
    \caption{Hiukkastiheys \SI{1}{\s} ja \SI{39}{\s} tarkennuksen jälkeen.} 
    \label{fig:unfiltered}
\end{figure}

Kuvassa \ref{fig:unfiltered} näkyy potentiaalinen ongelma: välittömästi tarkennuksen jälkeen simulaatio näyttää täsmälleen samalta. 
Simulaatioon muodostuu pieniä sokkeja, kun jakauman jyrkästi muuttuvilla alueilla on tasaisia kohtia. 
Tämä johtuu siitä, että tarkentaessa data kopioidaan suoraan tytärsoluille. 
Hienompi rakenne tulee esille simulaation jatkuessa.

\begin{figure}[p]
    \centering
    \includegraphics[width=0.9\textwidth]{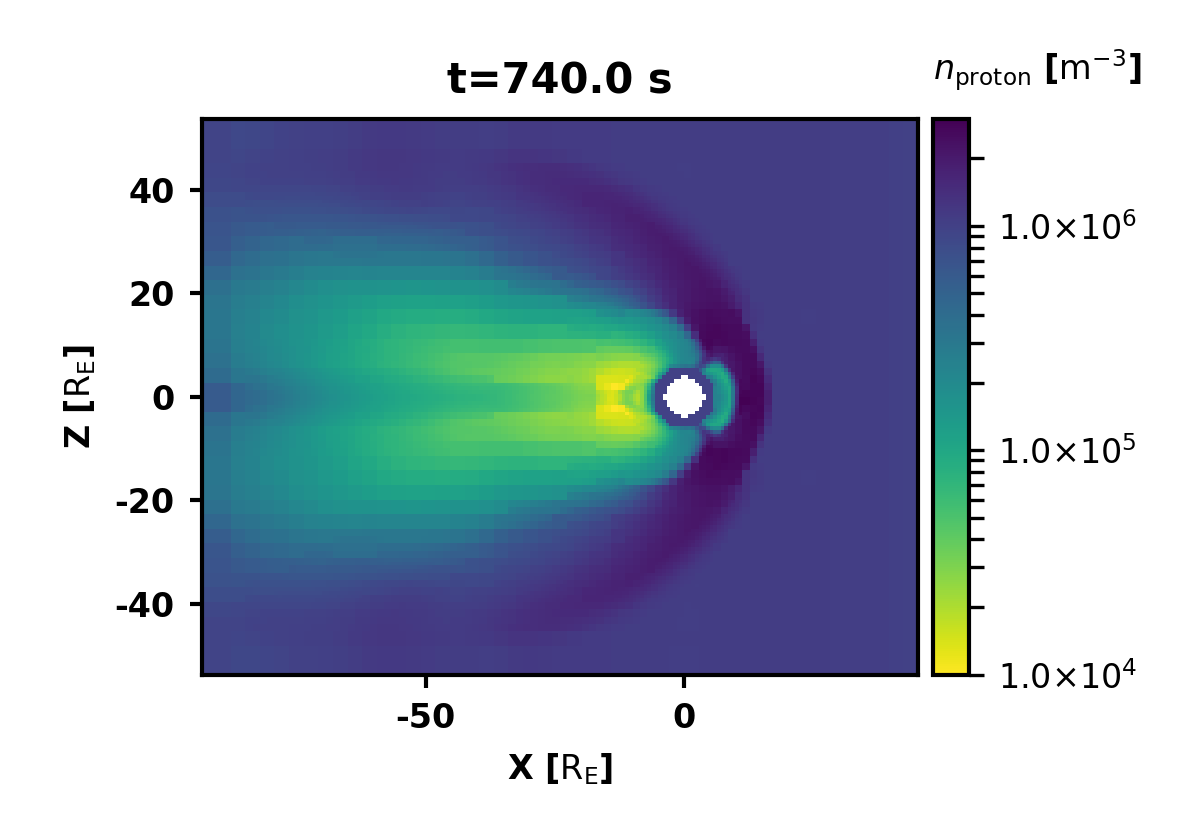}
    \includegraphics[width=0.9\textwidth]{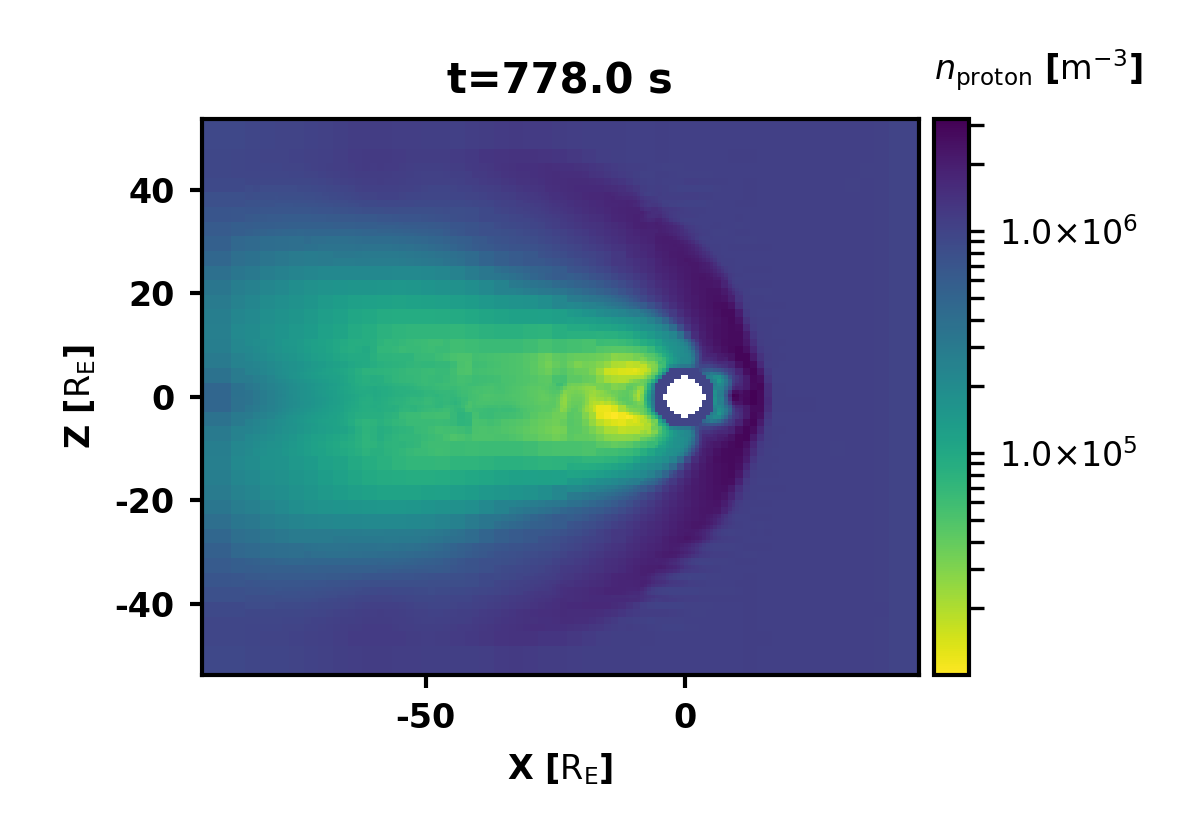}
    \caption{Hiukkastiheys \SI{1}{\s} ja \SI{39}{\s} tarkennuksen ja suodatuksen jälkeen.} 
    \label{fig:filtered}
\end{figure}

Edellämainittua ongelmaa voidaan korjata suodattamalla data tarkennuksen jälkeen.
Kuvassa \ref{fig:filtered} on käytetty liukuvaa keskiarvoa tarkennetuissa soluissa datan suodatukseen.
Etenkin keulasokin reuna näyttää tasaisemmalta jo heti tarkennuksen jälkeen.

\chapter{Johtopäätökset} \label{ch:johtopaatokset}
AMR on lupaava menetelmä kolmiulotteisessa mallinnuksessa.
Testiajossa huomataan selkeitä eroja aiempaan tarkennukseen, jotka ovat kuitenkin fysikaalisesti perusteltavissa: magnetosfäärin kärjen lisäksi koko keulasokki tarkentuu.
Ajonaikaisessa tarkennuksessa se ei kuitenkaan vielä toimi: simulaation hitaan tasoittumisen vuoksi tarkennusindeksi ei ainakaan heti tarkennuksen jälkeen ole käyttökelpoinen.
Datan suodattaminen auttaa, mutta sitäkään ei tule käyttää varomattomasti; liiallisella suodatuksella yksityiskohdat katoavat.

Mukautuva tarkennus soveltuu jo tällaisenaan paremmaksi heuristiikaksi simulaation tarkentamiseen.
Hienosäätämällä tarkennusindeksin kynnyksiä päästään toivottuun muistinkäyttöön. 

Menetelmää voidaan käyttää myös ajojen nopeampaan alustukseen.
Ajon alkaessa tyhjästä simulaatioalue alustetaan homogeeniseksi.
Aurinkotuulen virratessa sisään yhdestä suunnasta, vuorovaikuttaessa magnetosfäärin kanssa ja virratessa ulos simulaatio alkaa hitaasti muistuttaa todellista tilannetta.
Aloittamalla karkealla tarkkuudella tämä alkuvaihe voidaan ohittaa nopeammin;
kahdella tarkennustasolla tarkennettu ajo vei noin kymmenkertaisen määrän muistia tasaiseen verrattuna.

Kehitettävää kuitenkin on. Ajonaikainen tarkennus olisi toivottavaa sekä prosessin automatisoimiseksi että simulaation tarkkuuden parantamiseksi.
Testien perusteella suurin osa tarkennukseen kuluvasta ajasta tulee laskentatyön tasapainottamisesta prosessien välillä.
Tarkennus voitaisiin siis toteuttaa ennen normaalia tasapainotusta ilman merkittävää aikakustannusta.
Näiden tulosten perusteella menetelmän kehittäminen on perusteltua.


\begin{appendices}
\myappendixtitle

\chapter{Ajoparametrit} \label{ch:parameters}
\begin{lstlisting}
project = Magnetosphere
dynamic_timestep = 1
ParticlePopulations = proton

[restart]

[AMR]
max_spatial_level = 3
should_refine = 0
adapt_refinement = 0
refine_treshold = 1.0

[io]
diagnostic_write_interval = 1
write_initial_state = 0
restart_walltime_interval = 21000
write_restart_stripe_factor = 6
number_of_restarts = 12
vlsv_buffer_size = 0

system_write_t_interval = 1
system_write_file_name = bulk
system_write_distribution_stride = 0
system_write_distribution_xline_stride = 0
system_write_distribution_yline_stride = 0
system_write_distribution_zline_stride = 0

[gridbuilder]
x_length = 51
y_length = 40
z_length = 40
x_min = -6.12e8
x_max = 3.06e8
y_min = -3.6e8
y_max = 3.6e8
z_min = -3.6e8
z_max = 3.6e8

t_max = 2500.0

[proton_properties]
mass = 1
mass_units = PROTON
charge = 1

[proton_vspace]
vx_min = -4.0e6
vx_max = +4.0e6
vy_min = -4.0e6
vy_max = +4.0e6
vz_min = -4.0e6
vz_max = +4.0e6
vx_length = 20
vy_length = 20
vz_length = 20

[proton_sparse]
minValue = 1.0e-15
dynamicAlgorithm = 0
dynamicBulkValue1 = 1.0e6
dynamicBulkValue2 = 1.0e7
dynamicMinValue1 = 1.0e-15
dynamicMinValue2 = 1.0e-13

[Magnetosphere]
constBgBX = 0.0
constBgBY = 0.0
constBgBZ = -5.0e-9
noDipoleInSW = 1.0

dipoleType = 4
dipoleTiltPhi = 0.0 
dipoleTiltTheta = 0 
dipoleXFull = 9.5565e7 # 15 RE 
dipoleXZero = 2.5e8 
dipoleInflowBX = 0.0 
dipoleInflowBY = 0.0 
dipoleInflowBZ = 0.0 #-5.0e-9 

refine_L2radius = 11e7 # 17.27 RE
refine_L2tailthick = 2.5e7 # 3.92 RE
refine_L1radius = 1.59275e8 # 25 RE
refine_L1tailthick = 6.371e7 # 10 RE
refine_L3radius = 6.871e7 # 10.785 RE
refine_L3nosexmin = 5.0e7
refine_L3tailheight = 1.0e7
refine_L3tailwidth = 8.0e7
refine_L3tailxmin = -30.0e7
refine_L3tailxmax = -5.0e7

[ionosphere]
centerX = 0.0
centerY = 0.0
centerZ = 0.0
radius = 38.1e6
precedence = 2
reapplyUponRestart = 1

[proton_Magnetosphere]
T = 0.5e6
rho = 1.0e6
VX0 = -7.5e5
VY0 = 0.0
VZ0 = 0.0

nSpaceSamples = 1
nVelocitySamples = 1

[proton_ionosphere]
rho = 1.0e6
VX0 = 0.0
VY0 = 0.0
VZ0 = 0.0
taperRadius = 1e8

[loadBalance]
algorithm = RCB
rebalanceInterval = 50
tolerance = 1.2

[variables]
output = vg_rhom
output = vg_rhoq
output = vg_v
output = populations_vg_rho
output = populations_vg_v
output = populations_vg_moments_nonthermal
output = populations_vg_moments_thermal
output = populations_vg_acceleration_subcycles
output = fg_b
output = vg_b_vol
output = fg_e
output = fg_e_hall
output = vg_b_perturbed_vol
output = populations_vg_ptensor
output = vg_boundarytype
output = vg_boundarylayer
output = vg_rank
output = fg_rank
output = vg_loadbalance_weight
output = vg_maxdt_acceleration
output = vg_maxdt_translation
output = fg_maxdt_fieldsolver
output = populations_vg_blocks
output = vg_f_saved
output = populations_vg_effectivesparsitythreshold
diagnostic = populations_vg_blocks

output = vg_e_vol
output = vg_amr_drho
output = vg_amr_du
output = vg_amr_dpsq
output = vg_amr_dbsq
output = vg_amr_db
output = vg_amr_alpha
output = vg_amr_reflevel
output = vg_gridcoordinates

[proton_energydensity]
solarwindspeed = 7.5e5

[proton_precipitation]
nChannels = 9
emin = 500     # These are eV
emax = 50000

[boundaries]
periodic_x = no
periodic_y = no
periodic_z = no
boundary = Outflow
boundary = Maxwellian
boundary = Ionosphere

[outflow]
precedence = 3

[proton_outflow]
face = x-
face = y-
face = y+
face = z-
face = z+

[maxwellian]
face = x+
precedence = 4
reapplyUponRestart = 1

[proton_maxwellian]
dynamic = 0
file_x+ = sw1.dat

[bailout]
max_memory = 58
min_dt = 0.005

[fieldsolver]
maxSubcycles = 50
ohmHallTerm = 2
minCFL = 0.4
maxCFL = 0.45
maxWaveVelocity = 7494811.45  #2.5% of speed of light...    

[vlasovsolver]
minCFL = 0.8
maxCFL = 0.99
maxSlAccelerationRotation = 22
maxSlAccelerationSubcycles = 2
\end{lstlisting}
\begin{lstlisting}
0.0 1.0e6 0.5e6 -7.5e5 0.0 0.0 0 0 0
\end{lstlisting}

\end{appendices}


\cleardoublepage 
\phantomsection

\addcontentsline{toc}{chapter}{\bibname} 
\printbibliography

\end{document}